\documentclass[fleqn,usenatbib,useAMS]{mnras}

\usepackage{subfiles}
\usepackage{amsmath, graphicx, amssymb,xifthen,lipsum,comment,xcolor}

\newcommand{\changes}[1]{}
\newcommand{\deleted}[1]{}

\usepackage[T1]{fontenc}
\usepackage{ae,aecompl}

\usepackage{newtxtext,newtxmath}

\title[Plunging Region Electrons]{Nonthermal emission from the plunging region: a model for the high-energy tail of black hole X-ray binary soft states}

\author[A. M. Hankla et al.]{Amelia M. Hankla,$^{1, 2}$%
\thanks{E-mail: \href{mailto:lia.hankla@gmail.com}{lia.hankla@gmail.com}, \href{mailto:amelia.hankla@colorado.edu}{amelia.hankla@colorado.edu}}%
\thanks{National Science Foundation Graduate Research Fellow}
Nicolas Scepi,$^{1}$ Jason Dexter$^{1,3}$ 
\\
$^{1}$JILA, University of Colorado and National Institute of Standards and Technology, 440 UCB, Boulder, CO 80309-0440, USA\\
$^{2}$Department of Physics, University of Colorado, 390 UCB, Boulder, CO 80309-0390, USA\\
${^3}$Department of Astrophysical and Planetary Sciences, University of Colorado, 391 UCB, Boulder, CO 80309, USA}

\date{Last updated \today; in original form \today}

\pubyear{2022}

\begin{document}
\label{firstpage}
\pagerange{\pageref{firstpage}--\pageref{lastpage}}
\maketitle

\begin{abstract}
X-ray binaries exhibit a soft spectral state comprising thermal blackbody emission at 1 keV and a power-law tail above 10 keV. Empirical models fit the high-energy power-law tail to radiation from a nonthermal electron distribution, but the physical location of the nonthermal electrons and the reason for their power-law index and high-energy cut-off are still largely unknown. Here, we propose that the nonthermal electrons originate from within the black hole's innermost stable circular orbit (the ``plunging region''). Using an analytic model for the plunging region dynamics and electron distribution function properties from particle-in-cell simulations, we outline a steady-state model that can reproduce the observed spectral features. In particular, our model reproduces photon indices of $\Gamma\gtrsim2$ and power-law luminosities on the order of a few percent of the disk luminosity for strong magnetic fields, consistent with observations of the soft state. Because the emission originates so close to the black hole, we predict that the power-law luminosity should strongly depend on the system inclination angle and black hole spin. This model could be extended to the power-law tails observed above 400 keV in the hard state of X-ray binaries.
\end{abstract}



\section{Introduction} \label{sec:intro}
Observations of accreting black hole X-ray binary systems (XRBs) show two main X-ray spectral states. The hard state has an inverted power-law photon spectrum $N(E)\propto E^{-\Gamma}$ with a spectral index $\Gamma\lesssim2$ at 10 keV, peaking in the $100-200$ keV range. The soft state has a thermal blackbody component peaking around 1 keV and a high-energy power-law tail component from ~10 keV to 1 MeV. This hard X-ray to soft gamma-ray power law has a spectral index $\Gamma\gtrsim2$ and no observed cut-off~\citep{mcconnell2002,zdziarski2004,remillard2006}. 

Models of the soft spectral state usually involve a thin accretion disk that produces the thermal emission and a hot, optically-thin gas that produces the high-energy power-law tail. The thin accretion disk~\citep{ss73, novikovthorne} models the soft state's blackbody emission well. However, the origin of the power-law tail is less well-understood. Most models for the power-law tail assume a population of nonthermal electrons that are somehow accelerated to high energies, presumably by shocks or magnetic reconnection. The hybrid thermal/nonthermal electron distribution function fits the soft state spectrum up to 600 keV well~\citep{zdziarski2001,gierlinski1999}, but the electron distribution parameters are determined by the best fit to observational data rather than ab initio theory.

The spatial location of the hot, optically-thin gas and thus the nonthermal electrons is unknown. Hybrid electron distribution models usually assume that the nonthermal electrons originate from a jet or corona above a thin accretion disk extending down to the innermost stable circular orbit~\citep[ISCO;][]{zdziarski2004}. Alternatively, the nonthermal electrons could come from gas within the ISCO. In the ``plunging region,'' particle orbits transition from predominantly circular to predominantly radial. The ISCO serves as an artificial boundary condition in early thin disk models, which assume that the stress there goes to zero and no light is emitted from within the plunging region~\citep{ss73, novikovthorne}. Later thin disk models find that stress at the ISCO can lead to extra dissipation, which is often parameterized as an additional accretion efficiency $\Delta\epsilon$~\citep{agol2000,gammie1999}. Magnetohydrodynamic (MHD) and general relavitistic MHD (GRMHD) simulations have measured additional dissipation within the ISCO on the order of $\Delta\epsilon\gtrsim0.2$ for weakly magnetized disks and up to twice the Novikov-Thorne efficiency for strongly magnetized disks~\citep{armitage2001,shafee2008, noble2009, penna2010,avara2016}. This extra dissipation could come from magnetic reconnection and turbulence occurring in the plunging region.

Extra dissipation from within the plunging region could cause the soft state's high-energy power-law tail. In a purely thermal model, the gas temperature rises so rapidly towards the event horizon that the Wien peaks sum to a power law~\citep{zhu2012}. This model could also explain high-frequency quasi-periodic oscillations in the steep power law state~\citep{dexter2014}. These models assume that the electron and proton gas are in thermal equilibrium. However, shorter infall times in the plunging region can mean electrons and protons no longer have time to thermalize. Estimates of the electron-proton thermalization time in GRMHD simulations become shorter than the infall time close to the black hole, suggesting that electrons and protons may decouple into a two-temperature plasma within the ISCO~\citep{zhu2012}. 

In this work, we present a model for the soft spectral state's high-energy power-law tail using nonthermal electrons in the plunging region. We outline the parameter space where electron-proton decoupling should occur and demonstrate that the plunging region conditions are conducive to accelerating electrons to nonthermal energies (Section~\ref{sec:dynamical}). We then propose a model for the hybrid thermal/nonthermal electron distribution in the plunging region based on particle acceleration prescriptions from particle-in-cell simulations (Section~\ref{sec:methods}). We show that our model can reproduce observational characteristics of a high-energy power-law tail for highly-magnetized parameters, recover the transition to thermal-dominated spectra at lower magnetizations, and predict trends with inclination angle and black hole spin that are consistent with current observations (Section~\ref{sec:results}). We end by discussing our model assumptions and its general applicability to XRBs (Section~\ref{sec:discussion}) and summarizing our findings (Section~\ref{sec:conclusions}).

\section{Dynamical Model for the Plunging Region} \label{sec:dynamical}
Within the ISCO, the accretion flow cannot be described as a thin disk. Particle orbits transition from mostly circular outside the ISCO to mostly radial inside the ISCO. The fast infall times in the plunging region mean that the usual assumptions of a collisional, single-temperature plasma need to be revisited. In this section, we will motivate our treatment of the plunging region as a two-temperature, highly-magnetized region with the potential to continuously accelerate particles to nonthermal energies. 

We will use~\citet{gammie1999}'s plunging region model as a backdrop for our model.~\citet{gammie1999} provides radial profiles of the gas number density $n_0$, four-velocity $u^\mu$, and radial magnetic field strength $B_r$ that agree well with numerical simulations~\citep{penna2010}. The model assumes a one-dimensional, steady-state MHD inflow in the plunging region, parameterized by black hole spin $a$, mass $M$, and the $F_{\theta\phi}$ component of the Maxwell tensor, which reduces to $-r^2B_r$ at the midplane in the nonrelativistic limit. Throughout, we assume an Eddington factor $\eta=0.1=L_{\rm Edd}/(\dot M_{\rm Edd} c^2)$, where $L_{\rm Edd}$ and $\dot M_{\rm Edd}$ are the Eddington luminosity and accretion rate, respectively. We avoid numerical artefacts due to the boundary condition $v^r=0$ at the ISCO by starting our solutions at $r_{\rm ISCO} - 0.1r_g$. We scale the model to physical units using a black hole mass of $10 M_\odot$ and a density scale height $H/r=1$, the latter choice motivated by Section~\ref{ssec:decoupling}'s observation that ions will not collide with electrons in the plunging region. 

\begin{figure*}
    \centering
    \includegraphics[width=\textwidth]{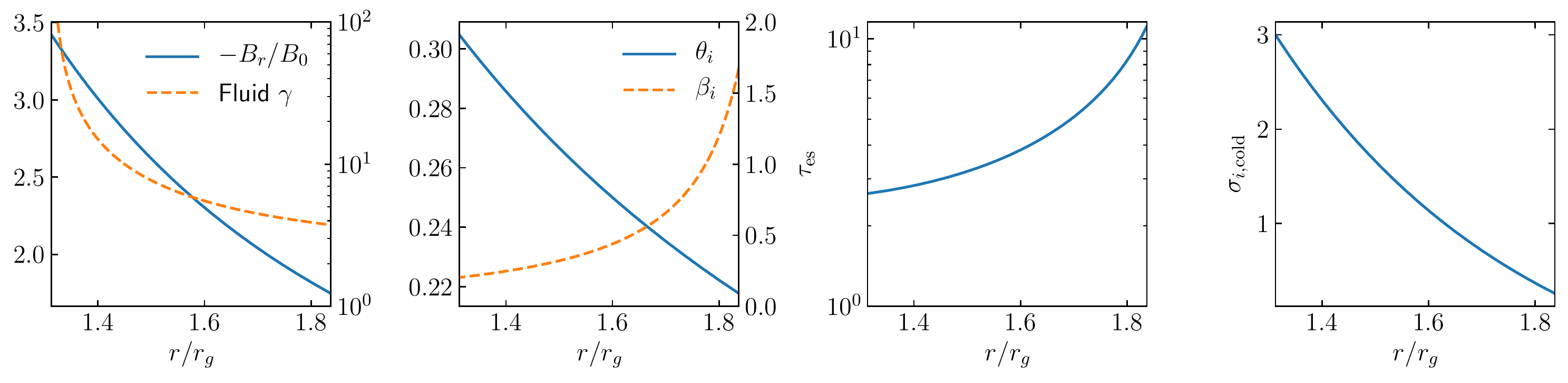}
    \caption{Properties of the plunging region calculated from the~\citet{gammie1999} background for fiducial parameters $a=0.95$ and $\Delta\epsilon=1.04$. From left to right: radial magnetic field strength $B_r$ (left axis) and the fluid's Lorentz factor (right axis), dimensionless ion temperature $\theta_i=k_BT_i/m_pc^2$ (left axis) and ratio of gas-to-magnetic pressure $\beta_i=8\pi n_0k_BT_i/B_r^2$ (right axis), electron scattering optical depth $\tau_{\rm es}$, and cold ion magnetization $\sigma_i$. Here, $B_0=10^8$ G. The plunging region extends from the ISCO at $1.94 r_g$ to the event horizon at $1.31 r_g$ in the midplane.}
    \label{fig:gammie_quantities}
\end{figure*}

For our fiducial model, we use values $a=0.95$ and $F_{\theta\phi}=6.0$, which corresponds to an accretion efficiency $\Delta\epsilon=1.04$. For all models, we assume that $\Delta\epsilon$ is uniformly distributed throughout the volume within the ISCO and that this heating comes from magnetic reconnection processes. Representative quantities of the fiducial background are plotted in Figure~\ref{fig:gammie_quantities}. For these parameters, the magnetic field has a strength on the order of $10^8$ G. We assume a virial ion temperature $T_i=2GMm_p/5k_Br$, which gives mildly relativistic ions, and calculate the ion beta $\beta_i=8\pi n_0k_BT_i/B_r^2$. The electron scattering optical depth $\tau_{\rm es}=n_0\sigma_T H$ is always slightly greater than 1. The values of $B_r$, $\tau$, and $\beta_i$ vary by a factor of $\sim 3$ from the ISCO to the event horizon. 

\subsection{Development of a Two-temperature Plasma}\label{ssec:decoupling}
Analytic models of a thin disk that permit non-zero stress at the ISCO predict a factor of ten increase in the gas temperature between the ISCO and the event horizon~\citep{agol2000}. Single-temperature GRMHD simulations find a similar increase \citep{zhu2012} due to the flow becoming effectively optically thin. The rapid rise of temperature means that the timescale for electrons and protons to reach thermal equilibrium increases as well. This electron-ion thermalization timescale $t_{\rm th}^{ei}$ is given by
\begin{equation}
    \frac{dT_e}{dt}=\frac{T_i-T_e}{t_{\rm th}^{ei}}
\end{equation}
where $t_{\rm th}^{ei}\equiv1/\nu_{\rm th}$. For hot protons heating electrons, the thermalization frequency from~\citet{stepney1983} reads:
\begin{align}
    \nu_{\rm th} &= \frac{m_e}{m_p}\frac{n_0\sigma_T c\ln\Lambda}{K_2(1/\theta_e)K_2(1/\theta_i)}\\
    &\left[\frac{2(\theta_e+\theta_i)^2+1}{\theta_e+\theta_i}K_1\left(\frac{\theta_e+\theta_i}{\theta_e\theta_i}\right)+2K_0\left(\frac{\theta_e+\theta_i}{\theta_e\theta_i}\right)\right]~{\rm s^{-1}}\\
    &= 3.6\times10^{-37}\frac{\left(m_em_p\right)^{1/2}n_0\ln\Lambda}{\left(m_pT_e+m_eT_i\right)^{3/2}}~{\rm s^{-1}},
\end{align}
where $\theta_e=k_BT_e/m_ec^2$ and $\theta_i=k_BT_i/m_pc^2$, $m_e$ and $m_p$ are electron and proton masses, respectively. The last line takes the nonrelativistic limit assuming $\theta_e\ll1$ and $\theta_i\ll1$~\citep{spitzer1956}. Here, $K_0(x)$, $K_1(x)$, and $K_2(x)$ are modified Bessel functions of the second kind. Throughout, we set the Coulomb logarithm $\ln\Lambda = 10$ and employ cgs units.

In the plunging region, the thermalization timescale can become longer than the accretion timescale.~\citet{zhu2012} finds that a weakly magnetized flow cannot equilibrate within $3~r_g$ for black hole spin $a=0.7$. Therefore, the assumption that electrons and protons couple efficiently no longer applies within the ISCO. The inefficient Coulomb coupling means that protons cannot cool, and should heat due to viscous and magnetic dissipation. Such increased temperature and thus increased ion pressure support will puff up the plunging region beyond a thin disk. The formation of a thick inner disk and a thin outer disk was recently simulated for the first time~\citep{liska2022}. For those model parameters, the thin outer disk with $H/r\sim10^{-2}$ expanded to $H/r\sim0.3$ within the ISCO.

This decoupling motivates our subsequent treatment of the electrons and protons in the plunging region as completely independent, with different temperatures, as well as our assumption that $H/r\sim1$. Indeed, Figure~\ref{fig:timescales} shows that in our model, the thermalization time (black solid line) is longer than the infall time (dashed horizontal line) for a wide range of accretion efficiencies. The same hierarchy holds for all black hole spins discussed in this work. 

\subsection{Nonthermal Particle Acceleration} \label{ssec:ntpa}
Particle-in-cell (PIC) simulations of magnetic reconnection have shown that efficient particle acceleration in an electron-ion plasma occurs for ion magnetizations $\sigma_i\gtrsim1$~\citep{werner2018}. Using background profiles for magnetic field and gas density from~\citet{gammie1999}, we will demonstrate that efficient particle acceleration is possible inside the ISCO. We calculate the ion magnetization
\begin{equation}
    \sigma_i=\frac{B_0^2}{4\pi n_0 m_pc^2}
\end{equation}
using the definition from~\citet{werner2018}. We set $B_0=B_r$ and $n_0=\rho/m_p$, with the radial magnetic field $B_r(r)$ and mass density $\rho(r)$ obtained from~\citet{gammie1999}. For fiducial model parameters, the ion magnetization ranges from 0.2 to 3.0 (Figure~\ref{fig:gammie_quantities}). In PIC simulations of magnetic reconnection with similar ion magnetizations,~\citet{werner2018} finds that electrons are accelerated into power-law distribution with high-energy cut-off $\gamma_c\sim4\sigma_e$ (rising slowly with $\sigma_i$) and power-law index $p$ prescribed by
\begin{equation}
    p=1.9+0.7/\sqrt{\sigma_i}.\label{eq:pprescription}
\end{equation}
\citet{werner2018} fixes $\theta_i=\sigma_i/200$ (i.e. $\beta_i=4\theta_i/\sigma_i\to0.02$), a factor of 10 to 100 times colder than the equipartition ion temperature calculated above. PIC studies of electron-ion turbulence (rather than magnetic reconnection) that set $\beta_i=2/3$ and scanned $\theta_i\in[1/256, 10]$ with $\sigma_i\in[1/256, 10]$ have found broadly similar power-law indices~\citep{zhdankin2019}. The turbulent simulations in general find more efficient particle acceleration, measuring $p\approx2.8$ for $\theta_i=1/16$ and $\sigma_i=0.2$ when Equation~\ref{eq:pprescription} would predict $p=3.5$. 

With ion magnetizations close to unity, the plunging region can plausibly accelerate electrons to nonthermal energies via turbulence or reconnection. This acceleration could happen in a two-stage process, with sharp increases in particle Lorentz factor when the particle crosses a current sheet and slower, second-order Fermi-like acceleration that continues to increase particle energy over time~\citep{comisso2019}. For a magnetically-dominated pair plasma, the current sheet acceleration time is on the order of $t_{\rm acc}\sim (300/\sigma)\omega_{pe}^{-1}$~\citep{comisso2019}, where $\omega_{pe}=\sqrt{4\pi e^2 n_0/m_e}$ is the electron plasma frequency. For the fiducial parameters $\Delta\epsilon=1.04$ and $a=0.95$, the~\citet{gammie1999} model yields $100\omega_{pe}^{-1}\sim 10^{-11}~{\rm s}$, much shorter than infall times on the order of $10^{-4}~{\rm s}$ for a ten solar-mass black hole and electron-ion thermalization times (Figure~\ref{fig:timescales} orange line). Therefore, we argue that the plunging region can accelerate electrons locally, forming a power-law distribution that will not have time to be thermalized by protons.

\subsection{Electron-electron collisions}\label{ssec:eecollisions}
A pure power-law distribution of electrons will interact via Coulomb collisions, eventually forming a pure thermal distribution if allowed to evolve freely~\citep{nayakshin1998}. Lower-energy electrons thermalize first, forming a hybrid thermal/nonthermal population at intermediate evolution times. If nonthermal electrons are continuously injected, as Section~\ref{ssec:ntpa} showed is possible in the plunging region, the electron distribution will reach some equilibrium of nonthermal and thermal populations. We model this hybrid distribution as a thermal Maxwell-J\"uttner distribution up to a transition Lorentz factor $\gamma_1$, where the distribution becomes a power law. A nonthermal electron with Lorentz factor $\gamma$ will interact with the thermal bulk on the energy loss timescale $t_{\rm coll}^{ee}(\gamma)\equiv1/\nu_\epsilon^{ee}(\gamma)$. The energy loss rate is given by
\begin{equation}
    \nu_\epsilon^{ee}(\gamma)=\left[\psi(x^{ee})-\psi'(x^{ee})\right]\frac{16\sqrt{\pi} e^4\ln\Lambda~n_0}{m_e^2c^3\beta^3}, \label{eq:nucoll}
\end{equation}
where $\beta=\sqrt{1-1/\gamma^2}$ and $\psi(x)$ is the lower incomplete Gamma function~\citep{spitzer1956}. The kinetic ratio $x^{ee}=\beta^2/(2\theta_e)$. For an electron temperature of $10^9$ K, a particle with a Lorentz factor of 2.0 has an energy loss timescale of $10^{-6}$ s, about 0.002 times the infall timescale (Figure~\ref{fig:timescales} purple). We therefore cannot neglect electron-electron collisions, especially at lower electron energies. 

Though nonthermal electrons will also collide with thermal protons over the corresponding electron-proton energy loss timescale, this timescale will be larger than the thermalization timescale discussed in Section~\ref{ssec:decoupling}. We will therefore ignore interactions between nonthermal electrons and thermal ions. 

\subsection{Importance of Radiative Cooling} \label{ssec:cooling}
High-energy electrons cool rapidly due to either synchrotron or inverse Compton cooling. In our model, the cooling time $t_{\rm cool}(\gamma)$ is set by either synchrotron or inverse Compton cooling, depending on the ratio $U_B/U_{ph}$, where $U_B=B^2_r/8\pi$ is the magnetic energy density and $U_{ph}=L/(4\pi c~r_{\rm ISCO}^2)$ is the energy density in the ambient photons, assumed to come from the disk such that $L=0.1~L_{\rm Edd}$, where $L_{\rm Edd}=4\pi GM m_pc/\sigma_T=1.26\times10^{39}~{\rm erg/s}$. If $U_B/U_{ph}>1$, then the cooling time is set to the time it takes for an electron to lose half its energy due to synchrotron radiation~\citep{rybickilightman}:
\begin{equation}
    t_{\rm sync}(\gamma)=\frac{5.1\times10^8}{\gamma B_r^2}. \label{eq:tsync}
\end{equation}
In using the above equation, we assume that the particle is moving in gyro-orbits around the dominant magnetic field $B_r$, with a pitch-angle $\alpha=\pi/2$. If on the other hand $U_B/U_{ph}<1$, the cooling time is set to the inverse Compton cooling time, assumed to take the form
\begin{equation}
    t_{\rm IC}(\gamma)=\frac{U_B}{U_{ph}}t_{\rm sync}(\gamma). \label{eq:tIC}
\end{equation}
Notice that Equation~\ref{eq:tsync} and~\ref{eq:tIC} are valid only for ultra-relativistic particles. As such, we ignore cooling for electrons below a minimum Lorentz factor $\gamma_{\rm min}=2$, since below this Lorentz factor the cooling rate decreases rapidly.

\begin{figure}
    \centering
    \includegraphics[width=0.45\textwidth]{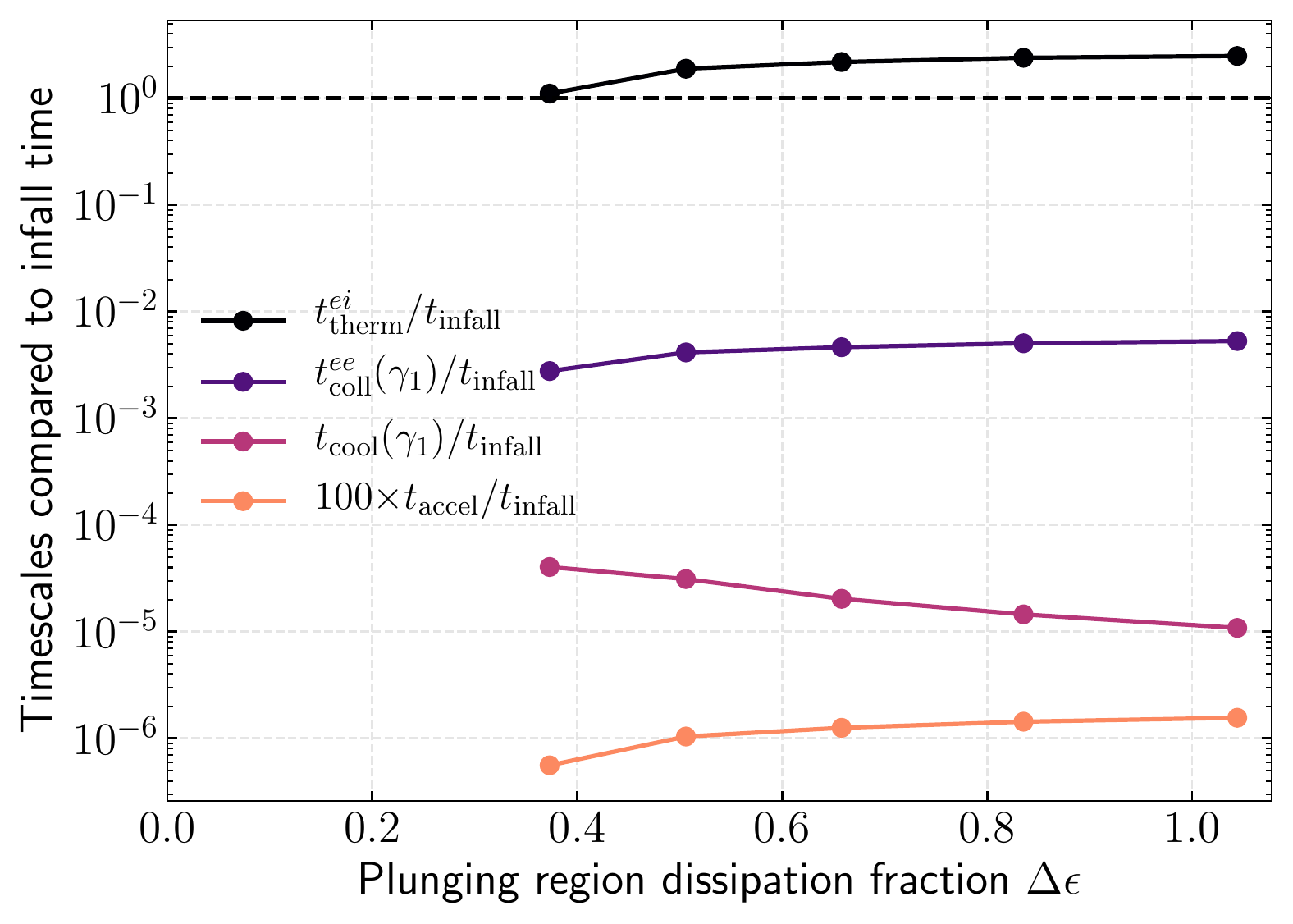}
    \caption{Timescales in the plunging region show that all physical processes except for electron-ion collisions are fast compared to the infall time for a wide range of accretion efficiencies $\Delta\epsilon$, consistent with the hierarchy $t_{\rm accel}\ll t_{\rm cool}(\gamma)\ll t_{\rm coll}^{ee}(\gamma)\ll t_{\rm infall} \lesssim t_{\rm therm}^{ei}$. Timescales are shown at a radius $r=1.58r_g$ for black hole spin $a=0.95$. These timescales are calculated for thermal ion and hybrid electron distributions determined by the steady-state model described in Section~\ref{sec:methods}.}
    \label{fig:timescales}
\end{figure}

The cooling times decrease for larger Lorentz factors, with values at the minimum Lorentz factor two orders of magnitude smaller than the electron-electron collision time, and five orders of magnitude smaller than the infall timescale (Figure~\ref{fig:timescales} magenta). The electrons in the plunging region therefore have time to radiate before disappearing beyond the event horizon.  

We show the relevant timescales of the plunging region in Figure~\ref{fig:timescales} for models with different black hole spin $a$ and accretion efficiencies $\Delta\epsilon$ resulting from different magnetic field strengths. At a radius of $r=1.58~r_g$, all models satisfy the ordering 
\begin{equation}
    t_{\rm accel}\ll t_{\rm cool}(\gamma)\ll t_{\rm coll}^{ee}(\gamma)\ll t_{\rm infall} \lesssim t_{\rm therm}^{ei}\label{eq:hierarchy}
\end{equation}
for $\gamma\ge\gamma_1$. We will use this ordering to motivate our model for the plunging region's hybrid thermal/nonthermal electron distribution.

\section{Multizone Equilibrium Model}\label{sec:methods}
In this section, we present a model for the electron distribution in the plunging region, motivated by the properties outlined in Section~\ref{sec:dynamical}. We will assume that nonthermal electrons are continuously injected by local processes such as magnetic reconnection or turbulence, which then partially thermalize via electron-electron collisions. We further assume that the hybrid thermal/nonthermal electron distribution reaches a steady state at each radius, matching the losses due to radiation with the heating due to magnetic dissipation. Below, we describe the assumptions that allow us to fully constrain this hybrid electron distribution's form.

\subsection{Initial Electron Distribution}
\begin{figure*}
    \centering
    \includegraphics[width=\textwidth]{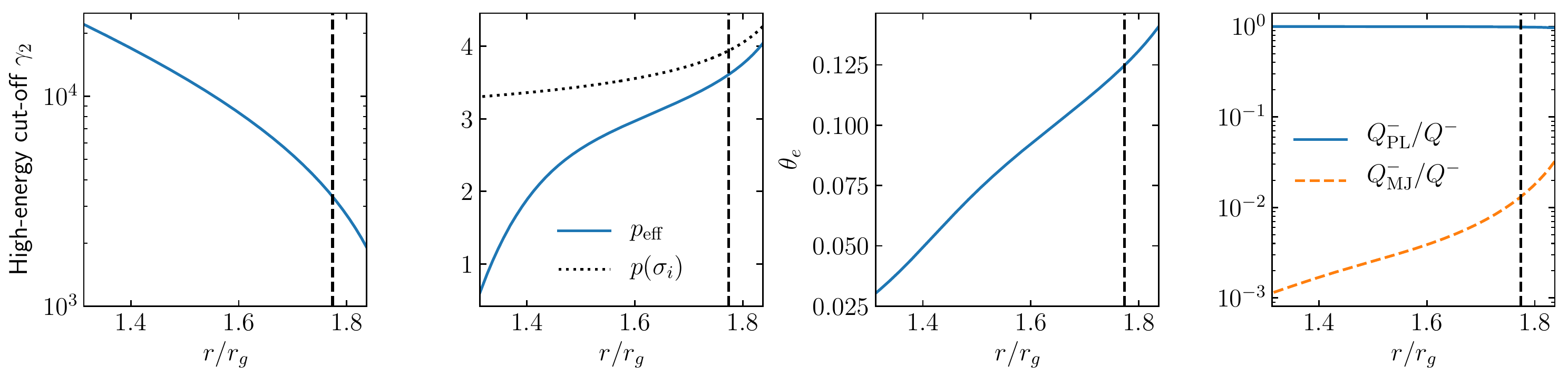}
    \caption{Radial profiles of the electron distribution function model properties for fiducial parameters $a=0.95$ and $\Delta\epsilon=1.04$. From left to right: the high-energy Lorentz factor cut-off of the power law $\gamma_2$, the electron power-law index, the thermal electron temperature $\theta_e=k_BT_e/m_ec^2$, and the fraction of the total cooling rate $Q_-$ from the power-law and thermal component. Vertical dashed line shows the location of the half-light radius.}
    \label{fig:prescription_quantities}
\end{figure*}

Because the advection timescales are much longer than the particle acceleration timescales (Figure~\ref{fig:timescales}), we assume that the initial (pre-thermalization) electron population at each radius is a pure power-law distribution: $f_{\rm PL,0}(\gamma)=A_0\gamma^{-p}$ for $1\le\gamma\le\gamma_2$, where $\gamma_2$ is a high-energy cut-off. The high-energy cut-off at each radius is set to
\begin{equation}
    \gamma_2(r) = \gamma_c(r) = 4\sigma_e(r), \label{eq:gamma2}
\end{equation}
where $\sigma_e=B_r^2(r)/(4\pi n_0m_ec^2)$ is the (cold) electron magnetization~\citep{werner2018}. The power-law index $p$ is also set by~\citet{werner2018}'s prescription (Equation~\ref{eq:pprescription}), increased by 1 to account for cooling~\citep{blumenthal1970}. Requiring the distribution to be normalized to $n_0$ sets the constant $A_0$. All together, the initial distribution reads:
\begin{align}
    f_{\rm PL,0}(\gamma)&=\frac{n_0(p-1)}{1-\gamma_2^{1-p}}~\gamma^{-p} & 1\le\gamma\le\gamma_2.\label{eq:fpl0}
\end{align}
Electron-electron Coulomb collisions will partially thermalize this initial power-law distribution. Radial profiles for the high-energy cut-off $\gamma_2$ and the power-law index $p$ in the fiducial model are given in Figure~\ref{fig:prescription_quantities}. 

\subsection{Steady-state Electron Distribution}
Some of the injected power-law electrons will interact with each other via Coulomb collisions, creating a thermal distribution below some Lorentz factor $\gamma_1$ (Section~\ref{ssec:eecollisions}). The resulting hybrid thermal/nonthermal electron distribution continuously gains more nonthermal electrons due to the injection process and loses nonthermal electrons to the thermal bulk, radiating energy. We approximate these rapid processes with a steady-state electron distribution comprising a Maxwell-J\"uttner thermal distribution $f_{\rm MJ}$ and a power-law distribution $f_{\rm PL}(\gamma)$ for $\gamma_1\le\gamma\le\gamma_2$. 

The thermal electron distribution $f_{\rm MJ}$ is fully described by two variables: its normalization and its temperature $T_e$. We fix the normalization of the thermal distribution to the number density $n_0$, assuming that the number of nonthermal electrons is negligible compared to the thermal electrons. Explicitly, the thermal distribution reads
\begin{equation}
    f_{\rm MJ}(\gamma,~\theta_e,~n_0)=n_0\frac{\gamma^2\beta}{\theta_eK_2(1/\theta_e)}e^{-\gamma/\theta_e}, \label{eq:fMJ}
\end{equation}
where $K_2$ is the modified Bessel function of the second kind. The electron temperature $T_e=\theta_e m_ec^2/k_B$ is set by the model assumptions outlined below.

The nonthermal electrons are described as a power-law between a minimum Lorentz factor $\gamma_1$ and a high-energy cut-off $\gamma_2$, with a power-law index $p_{\rm eff}$ and a normalization $A_{\rm PL}$:
\begin{align}
    f_{\rm PL}(\gamma)&=A_{\rm PL}\gamma^{-p_{\rm eff}} & \gamma_1\le\gamma\le\gamma_2. \label{eq:fPLform}
\end{align}
We will now outline how our model determines the variables $\gamma_1$, $A_{\rm PL}$, $p_{\rm eff}$, and $T_e$.

\subsubsection{Determining $\gamma_1$}
The minimum Lorentz factor $\gamma_1$ of the nonthermal distribution is set at the energy where collisions and cooling balance. At $\gamma_1$, the energy loss timescale $t_{\rm coll}^{ee}$ (Equation~\ref{eq:nucoll}) equals the cooling time $t_{\rm cool}$:
\begin{equation}
    t_{\rm coll}^{ee}(\gamma_1)=t_{\rm cool}(\gamma_1). \label{eq:gamma1eq}
\end{equation}
Here, the cooling time $t_{\rm cool}$ is set to the minimum of the synchrotron and inverse Compton cooling times (Equations~\ref{eq:tsync} and~\ref{eq:tIC}). We enforce the lower bound $\gamma_{1,\rm min}=\gamma_{\rm min}=2.0$ to model the inefficiency of cooling at low energies. We also require that $\gamma_1>\gamma_{\rm threshold}(T_e)$, where the threshold Lorentz factor $\gamma_{\rm threshold}$ is where the energy loss frequency (Equation~\ref{eq:nucoll}) is zero for a given temperature. A power-law electron with a Lorentz factor below $\gamma_{\rm threshold}$ would increase in energy due to collisions with the thermal bulk. Imposing the requirement that the power-law start at Lorentz factors greater than the threshold Lorentz factor means that $\gamma_1$ will always be to the right of the thermal peak, thereby avoiding getting trapped at solutions with $\gamma_1\to1$.

Often, the cooling times are shorter than the energy loss times for all values of Lorentz factor. In this case, collisions are never important for the nonthermal electrons (Figure~\ref{fig:timescales}). However, it is still important to maintain a thermal population of electrons to provide the bulk of the density in the plunging region. When no solution to Equation~\ref{eq:gamma1eq} exists, we set $\gamma_1=\max\left(\gamma_{1, {\rm threshold}}, \gamma_{\rm min}\right)$.

\subsubsection{Determining $A_{\rm PL}$}
The power-law normalization $A_{\rm PL}$ is set by requiring continuity of the thermal and nonthermal electrons at $\gamma_1$:
\begin{equation}
    f_{\rm MJ}(\gamma_1)=f_{\rm PL}(\gamma_1). \label{eq:continuity}
\end{equation}

\subsubsection{Determining $p_{\rm eff}$}
\begin{figure}
    \centering
    \includegraphics[width=0.45\textwidth]{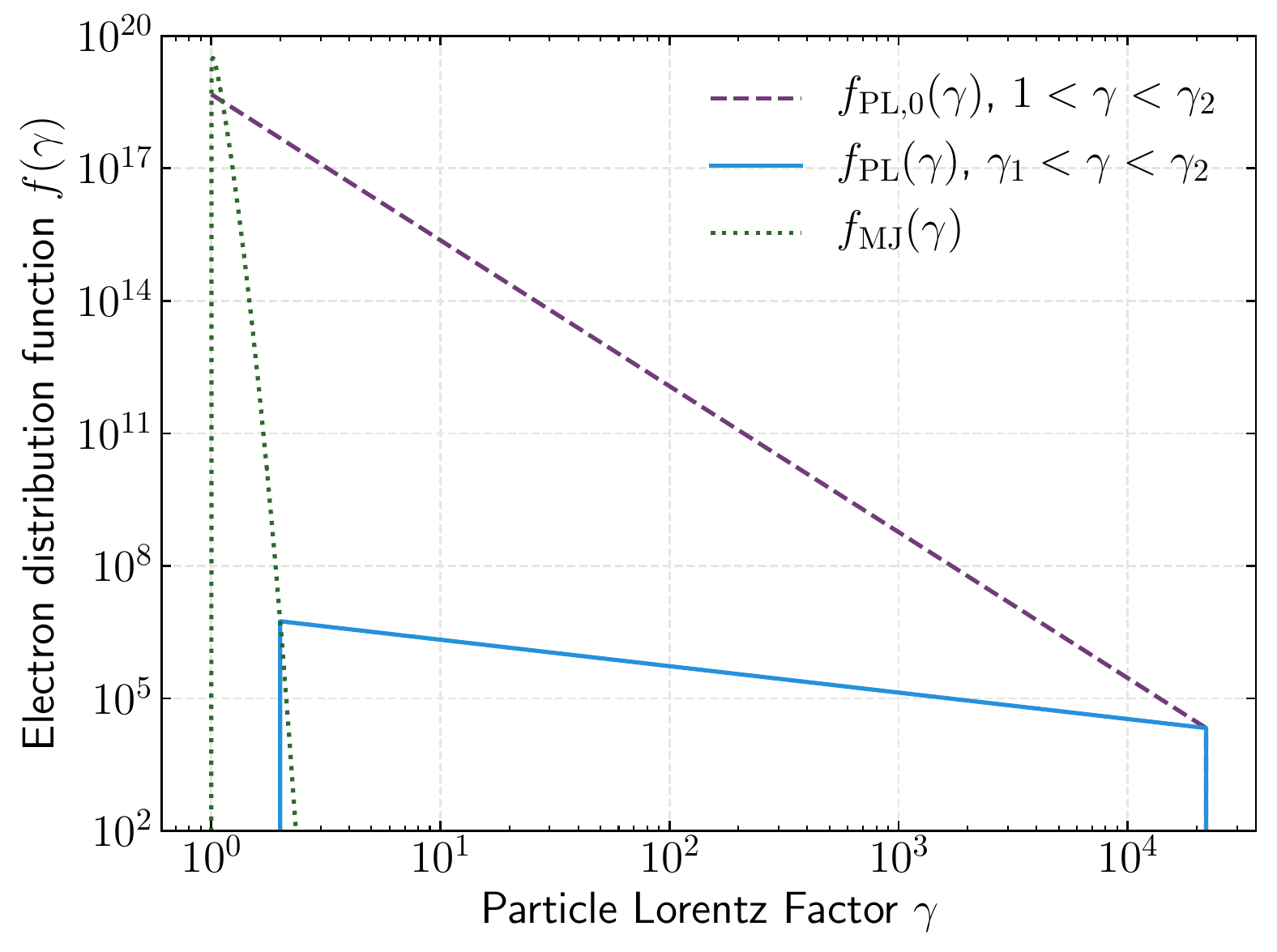}
    \caption{The effective power-law index $p_{\rm eff}$ is calculated from the initial power-law distribution $f_{\rm PL,0}$ by assuming that both power laws have the same value at the high-energy cut-off $\gamma_2$ (Equation~\ref{eq:peffeqgen}). This sample electron distribution was calculated for fiducial parameters close to the event horizon at a radius $r=1.31~r_g$, where $p=3.3$ (purple dashed line) and $p_{\rm eff}=0.6$ (blue solid line). For radii closer to the ISCO, the difference between $f_{\rm PL,0}$ and $f_{\rm PL}$ is not so pronounced. The thermal distribution is also shown for reference (green dotted line).}
    \label{fig:peffExample}
\end{figure}

The power-law index $p_{\rm eff}$ of the steady-state nonthermal distribution is calculated by assuming that none of the highest-energy particles are thermalized by Coulomb collisions; that is,
\begin{align}
    f_{\rm PL,0}(\gamma_2)=f_{\rm PL}(\gamma_2).\label{eq:peffeqgen}
\end{align}
We use an effective power-law index because high-energy particles collide more slowly than lower-energy particles, resulting in a power-law index that decreases over time~\citep{nayakshin1998}. Using the power-law index from PIC simulations would underestimate the total power law. Our formulation assumes that the initial power-law is continually replenished by in-situ particle acceleration, so the impact of cooling is primarily in the radiation rather than the dynamics. The difference between $p_{\rm eff}$ and $p$ is largest close to the event horizon  (Figure~\ref{fig:prescription_quantities}). Figure~\ref{fig:peffExample} shows how the final power-law distribution (blue solid line) differs from the initial power-law distribution (purple dashed line). 

Combining Eqns.~\ref{eq:fpl0} and~\ref{eq:peffeqgen} yields an expression for $p_{\rm eff}$:
\begin{equation}
    p_{\rm eff} = \frac{\log\left[ f_{\rm PL,0}(\gamma_2)/f_{\rm MJ}(\gamma_1)\right]}{\log\left[\gamma_1/\gamma_2\right]}. \label{eq:peffeq}
\end{equation}

\subsubsection{Determining $T_e$}
The temperature $T_e$ of the thermal electrons is set by requiring a steady state that balances energy lost to radiation with energy dissipated by magnetic torques. We obtain the volume heating rate $Q_+$ from the accretion efficiency $\Delta\epsilon$ given by the~\citet{gammie1999} model, assuming that dissipation occurs evenly throughout the plunging region's volume: 
\begin{equation}
    Q_+=\delta_e\frac{\Delta\epsilon~L_{\rm Edd}}{V_{\rm PR}}, \label{eq:heatingRate}
\end{equation}
where $V_{\rm PR}=4\pi(r_{\rm ISCO}^3-r_{\rm EH}^3)/3$ is the (Newtonian) volume of the plunging region. The fraction of dissipated energy $\delta_e$ that goes into electrons is set at a constant value of 0.5, as motivated by PIC simulations of the energy partition in an electron-ion plasma with magnetizations close to 1~\citep{werner2018,zhdankin2019}. The remainder $1-\delta_e$ fraction of dissipated energy heats the ions, which do not radiate.

The total cooling rate $Q_-$ sums the bremsstrahlung cooling rate $Q_{-, \rm MJ}$ from the thermal electrons with the inverse Compton or synchrotron cooling rate from the nonthermal electrons. To account for additional scattering, the nonthermal cooling rate $Q_{-,\rm PL}$ is set to the nonthermal emitted power $\mathcal{P}_{-, \rm PL}$ decreased by a factor of $\tau_e=n_0\sigma_T H=n_0\sigma_T (H/r)r$, such that $Q_{-,\rm PL}=\mathcal{P}_{-,\rm PL}/\tau_e$. This additional factor estimates how many of the synchrotron/IC photons would escape the plunging region without additional scattering, which is important since $\tau_e\gtrsim1$. This choice is motivated by Monte Carlo simulations of an isotropic sphere of gas with photons packets initialized evenly throughout the sphere. These simulations show that about $1/\tau_e$ of the electrons escape with no scattering. The remaining $1-1/\tau_e$ fraction of photons will Compton scatter off of the thermal/nonthermal electrons, heat the gas, and re-energize nonthermal electrons. The equation for steady-state thus reads:
\begin{equation}
    Q+=Q_-=Q_{-, \rm MJ} + Q_{-, \rm PL}.\label{eq:steadystate}
\end{equation}

The nonthermal emissivity $\mathcal{P}_{-, \rm PL}$ is calculated by integrating the Larmor formula for power emitted per electron over frequency $\omega$, the power-law electron distribution $f_{\rm PL}(\gamma)$, and solid angle. In integrating over solid angle $\Omega$, we assume an isotropic distribution of pitch-angles $\alpha$. The resulting power lost per unit volume is given by:
\begin{align}
    \mathcal{P}_{-, \rm PL}^{\rm sync}&=\int~d\Omega~d\omega~d\gamma~ f_{\rm PL}(\gamma) \frac{2e^4B_r^2\gamma^2\beta^2\sin^2\alpha}{3m_e^2c^3}\\
    &=\frac{16\pi e^4 B_r^2}{9m_e^2c^3}\int_{\gamma_1}^{\gamma_2} (\gamma^2-1)A_{\rm PL}n_0 \gamma^{-p_{\rm eff}} d\gamma.
\end{align}
If inverse Compton dominates, the emissivity is set to $\mathcal{P}_{-,\rm PL}^{\rm IC}=(U_{ph}/U_B) \mathcal{P}_{-, \rm PL}^{\rm sync}$. 

We describe the cooling due to thermal particles in the plunging region as saturated Comptonized bremsstrahlung. Because the optical depth is of order 1 and the Compton $y$ parameter is much greater than 1, the bremsstrahlung is amplified by repeated inverse Compton scatterings. The resulting emissivity is
\begin{equation}
    Q_{-, \rm MJ}=A(n_0, T_e)\epsilon_{\rm ff}(n_0, T)
\end{equation}
where $A(n_0, T_e)=0.74\left[\ln(2.25/x_{coh})\right]^2$ is an approximate amplification factor due to repeated inverse Compton scattering, $x_{coh}=2.4\times10^{17}(m_e n_0)^{1/2}T_e^{-9/4}$ and $\epsilon_{ff}(n_0, T)=1.68\times10^{-27}T_e^{1/2}n_0^2$ is the standard bremsstrahlung emissivity~\citep{rybickilightman}. 

Because the energy loss collision times (Equation~\ref{eq:nucoll}) assume a nonrelativistic thermal bulk, we put an upper limit on $T_e$ at the temperature $T_{\rm max}$ where setting $\nu_\epsilon^{ee}(x^{ee}_c)=0$ would require a particle velocity greater than the speed of light. This maximum temperature is $T_{\rm max}=3\times10^9$ K.

Equations~\ref{eq:gamma1eq},~\ref{eq:peffeqgen},~\ref{eq:continuity}, and~\ref{eq:steadystate} are solved iteratively at each radius using Brent's method to obtain $\gamma_1(r)$, $p_{\rm eff}(r)$, $A_{\rm PL}(r)$, and $T_e(r)$. 

\section{Results} \label{sec:results}
\subsection{High-spin, highly-magnetized case}\label{ssec:fiducial}
To illustrate the model, we examine the fiducial case $a=0.95$ and $\Delta\epsilon=1.04$, which corresponds to $F_{\theta\phi}=6.0$ and magnetic flux a factor of ~2 below the saturation value for a magnetically-arrested disk~\citep{avara2016}. The hierarchy of timescales for this fiducial model justifies the assumptions outlined in Section~\ref{sec:dynamical}. For this set of parameters, $\gamma_1$ always hits the lower bound $\gamma_{1,\rm min}=2.0$. The equilibrium electron temperature $\theta_e$ decreases from 0.14 ($8.9\times10^8$ K) at the ISCO to 0.03 ($1.9\times 10^8$) K at the event horizon (Figure~\ref{fig:prescription_quantities}). These temperatures are roughly consistent with the inner disk temperatures in two-temperature GRMHD simulations~\citep{liska2022}. The temperature decreases towards the event horizon because of the increase in $\gamma_2$, decreasing the power law's overall normalization and shifting the temperature lower because of the continuity requirement (Equation~\ref{eq:continuity}). The decoupling of electrons and protons happens very close to the ISCO, at $1.82~r_g$, meaning that almost the entire plunging region has decoupled. The cooling rate is dominated by nonthermal synchrotron emission, with thermal cooling representing $\lesssim1\%$ of the cooling rate at the ISCO and $0.1\%$ at the event horizon (Figure~\ref{fig:prescription_quantities}). This increase in the fraction of cooling from nonthermal electrons is again driven by the increase in $\gamma_2$, which is set by the background magnetic field and gas density. 

\begin{figure}
    \centering
    \includegraphics[width=0.45\textwidth]{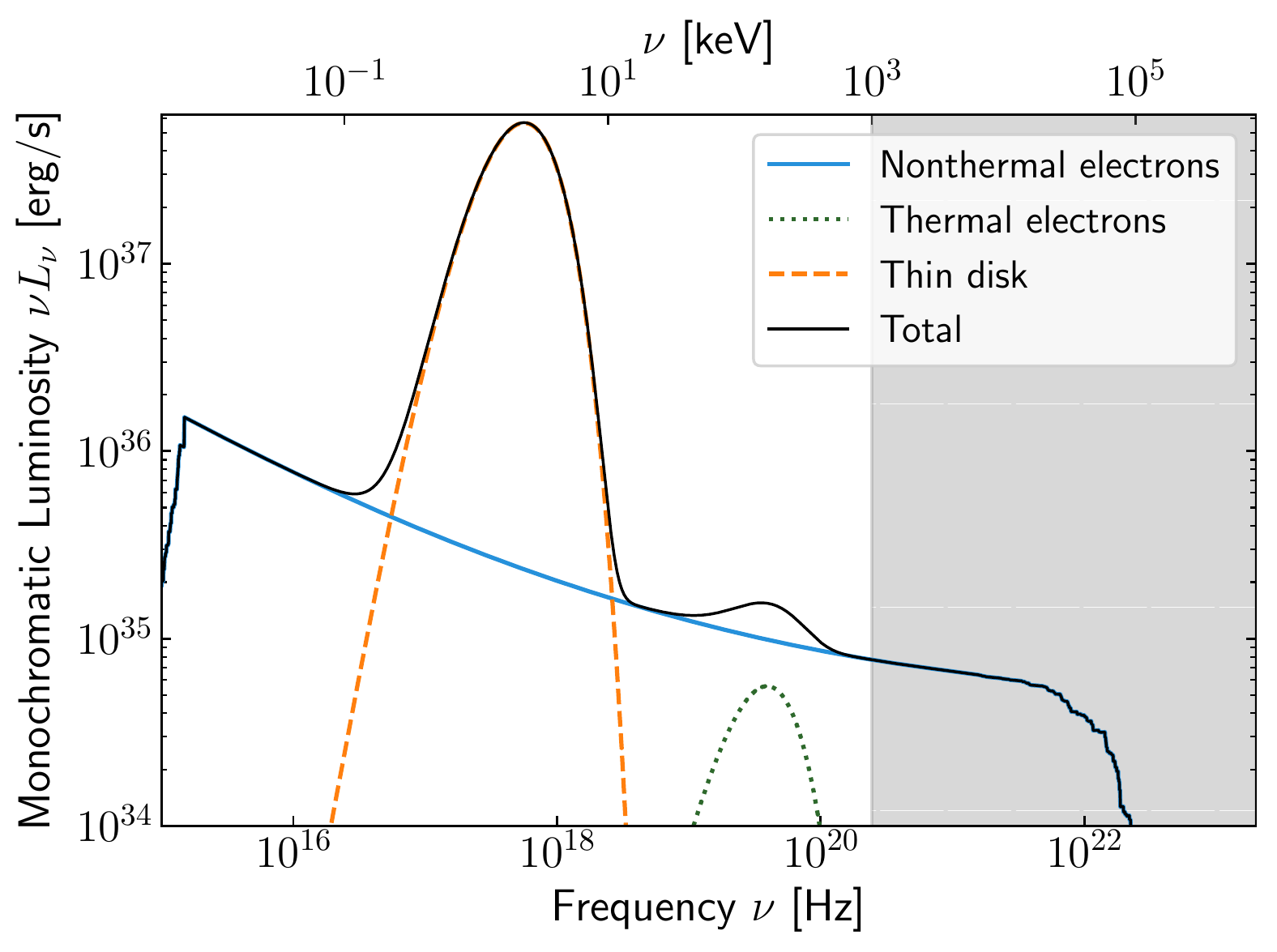}
    \caption{Spectrum for the fiducial model parameters $a=0.95$ and $F_{\theta\phi}=6.0$ at an inclination of $60^\circ$. The thin disk emission (dashed orange line) peaks around a few keV. The nonthermal electrons' radiation (solid blue line) extends from 10 eV to 0.1 GeV, although emission beyond 1 MeV (gray region) may be impacted by pair creation. The thermal electrons in the plunging region create a small excess at around 10 keV (dotted green line).}
    \label{fig:exspectrum}
\end{figure}

The total spectrum, shown in Figure~\ref{fig:exspectrum}, consists of three parts: the plunging region's nonthermal electrons (blue solid line), the plunging region's thermal electrons (green dotted line), and the thin disk's spectrum (orange dashed line). The spectrum is calculated with redshifts obtained by ray tracing the plunging region-disk system with the code \texttt{geokerr} to a camera inclined at $i=60^\circ$ to the black hole and disk angular momentum vector~\citep{dexter2009}. The gas is assumed to have a four-velocity given by the~\citet{gammie1999} model within the plunging region and a purely azimuthal, Keplerian velocity in the disk body. The power law intensities $I_{\nu, {\rm PL}, ij}=P_{\nu, {\rm PL},ij}H/(4\pi)$ at a camera pixel with position $i$ and $j$ in the $x-$ and $y-$directions are calculated by assuming a hybrid electron distribution at the radius $r_{ij}$ where the ray-traced photon hits the plunging region midplane. The power-law distribution is $P_{\nu, {\rm PL},ij}=A_{ij}\nu^{-s_{ij}}$ between the characteristic synchrotron frequencies $\nu_{1,ij}$ and $\nu_{2,ij}$ for electrons with Lorentz factors $\gamma_{1,ij}$ and $\gamma_{2,ij}$. That is, $\nu_1=\nu_0\gamma_{1,ij}^2$ and $\nu_2=\nu_0\gamma_{2,ij}^2$, where $\nu_0=3e|B_r|/(4\pi m_e c)$. The power-law slope is $s_{ij}=(p_{\rm eff,ij}-1)/2$. The constant $A_{ij}$ is set by requiring that $\int P_{\nu, {\rm PL},ij}d\nu=Q_{-,\rm PL,ij}$. The emission from the thermal plunging region electrons is in the Comptonized Wien regime, such that $I_{\nu, {\rm MJ}, ij}=I_\nu^W(T_{ij})e^{-\alpha_{ij}}$, where $T_{ij}$ is the thermal electron temperature at $r_{ij}$, $I_\nu^W(T_{ij})=2h\nu^3/c^2e^{-kT_{ij}/h\nu}$ is the Wien intensity, and $e^{-\alpha_{ij}}$ is the Compton saturation factor. The Compton saturation factor is determined by setting the frequency-integrated intensity equal to the isotropic thermal Comptonized bremsstrahlung emissivity: $I^W(T_{ij})e^{-\alpha_{ij}}=Q_{-,\rm MJ}H/4\pi$. Luminosities are obtained by summing over $64\times64$ camera pixels: $L_\nu=4\pi\Delta\alpha\Delta\beta\sum_{ij}I_{\nu,ij}$, where $\Delta\alpha=\Delta\beta=0.125r_g$ are the pixel widths. The disk spectrum is calculated by assuming a blackbody with a radially-dependent temperature extending from the ISCO to $10^{10}$ cm~\citep{FKR}: 
\begin{equation}
    T(r)=\left(\frac{3GM\dot M}{8\pi\sigma r^3}\frac{\mathscr{Q}}{\mathscr{B}\mathscr{C}^{1/2}}\right)^{1/4}
\end{equation}
where $\mathscr{B}$, $\mathscr{C}$, and $\mathscr{Q}$ are relativistic corrections~\citep{novikovthorne,pageThorne}.

Several features of the spectrum in Figure~\ref{fig:exspectrum} stand out. First, emission from the nonthermal plunging region electrons dominates the  emission above 10 keV and produces an observable high-energy power-law tail. The power law has a relatively flat slope past 200 keV, a feature that is difficult to produce through thermal Comptonization~\citep{coppi1999}. This power law cuts off at a little less than 100 MeV, though above 1 MeV pair production could become important (gray region). The total luminosity from the power law across all frequencies is 6.1\% of the disk luminosity. This fraction decreases to 1.7\% if non-X-ray frequencies below 1keV are excluded. Excluding luminosity from the pair-emitting frequency range above 1 MeV further decreases the fraction to 1.3\%. The nonthermal power-law electrons dominate emission from the plunging region, with a luminosity 63, 15, or 12 times that of the thermal plunging region electrons for all frequencies, frequencies above 1 keV, and frequencies between 1keV and 1 MeV, respectively.

\begin{figure}
    \centering
    \includegraphics[width=0.45\textwidth]{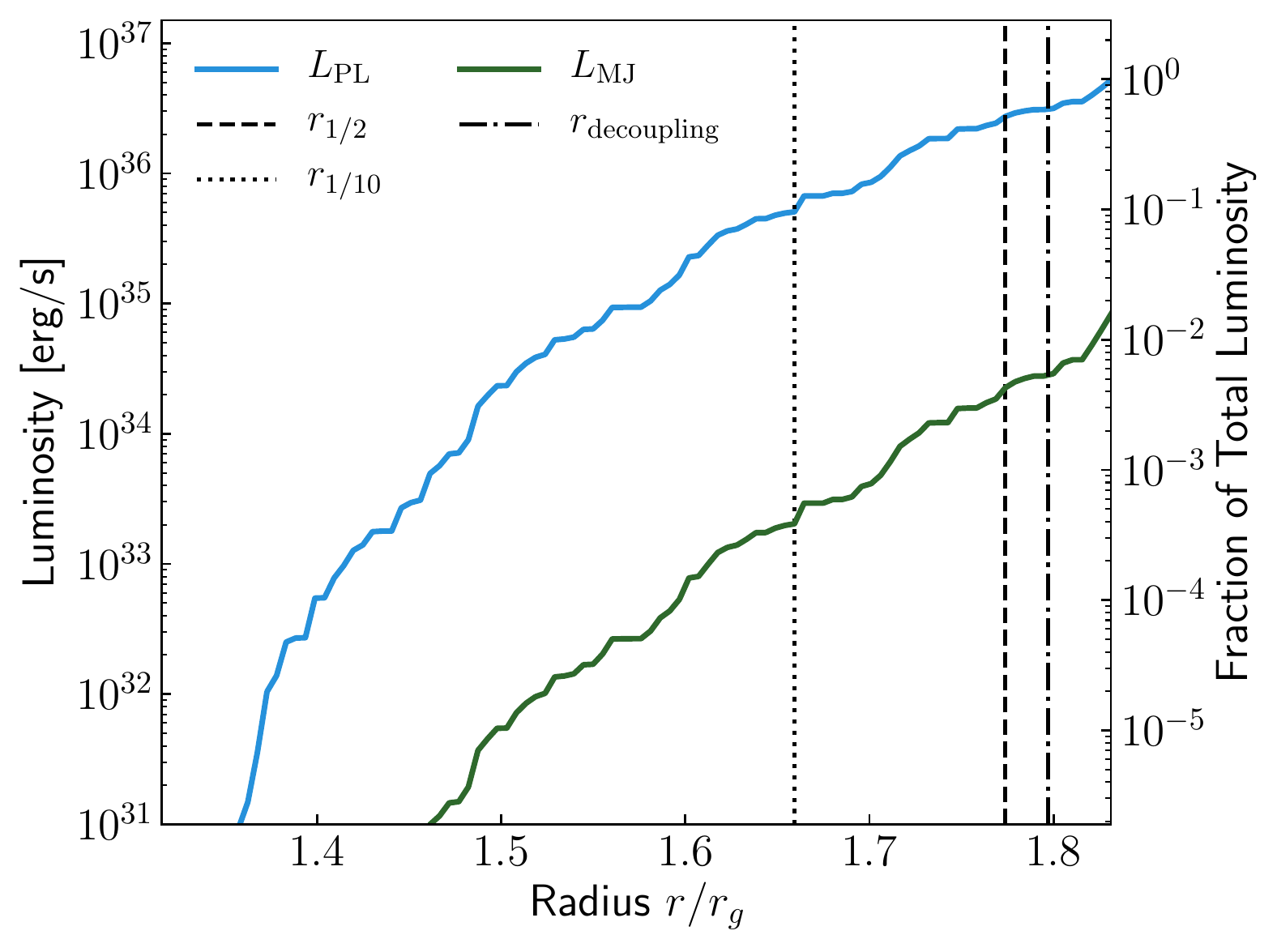}
    \caption{Portion of the fiducial model's observed frequency-integrated luminosity originating between the event horizon and a radius $r$ for an inclination angle of $60^\circ$. Emission from the plunging region's nonthermal electrons (blue) dominates over the plunging region's thermal electrons (green). 10\% of the luminosity comes from $r<r_{1/10}=1.66r_g$ (dotted line), while 50\% comes from $r<r_{1/2}=1.77r_g$ (dashed line).  Vertical dash-dot line shows the decoupling radius.}
    \label{fig:luminosity}
\end{figure}

Most of the observed luminosity comes from radii close to the ISCO, as shown in Figure~\ref{fig:luminosity}. This figure calculates the amount of observed luminosity that originates from between the event horizon and a radius $r$. We define the half-light radius $r_{1/2}$ as the location within which 50\% of the light from the plunging region has been emitted; analogously, $r_{1/10}$ is where 10\% of the light has been emitted. For this fiducial case with the ISCO at $r=1.94r_g$, 50\% of the light comes from $r>1.77r_g$ and 90\% comes from $r>1.66r_g$. 

The emission depends on the spin and the inclination of the disk-plunging region system. The inclination of the disk with respect to an observer changes the effective observed area, an effect that goes as $\cos i$, where $i$ is the angle between the black hole spin axis and the line-of-sight. Relativistic beaming effects have the opposite trend with inclination angle; radiation from relativistically-moving gas is beamed into the plane of motion, leading to a higher luminosity for larger inclination angles. In the plunging region, beaming effects dominate, resulting in a factor of 100 increase in power-law luminosity from $i=0^\circ$ ($L_{\rm PL}=5\times10^{35}$ erg/s) to $i=90^\circ$ ($L_{\rm PL}=6\times10^{37}$ erg/s) for $a=0.95$. Power-law emission for $a=0.5$ is roughly constant over inclination angle, with $L_{\rm PL}\sim 2-3\times10^{36}$ erg/s. Beaming effects no longer dominate for the smaller spin because the fluid Lorentz factor is smaller due to a larger $r_{\rm ISCO}$. The blackbody disk's peak temperature depends on the location of the ISCO, which is set by the black hole spin and decreases by a factor of about 3 from a non-spinning to a maximally-spinning black hole~\citep{cunningham1975, li2005}. The disk emission is non-monotonic as a function of inclination angle, peaking around $60^\circ$, where area projection effects start to dominate over Doppler effects. The peak disk emission is a factor of two larger than at $i=0^\circ$ for $a=0.95$, and a factor of 1.7 larger for $a=0.5$. As a result, $L_{\rm PL}/L_{\rm disk}({\rm 1keV}<\nu<{\rm 1MeV})$ for $a=0.95$ increases from 0.002 at $i=0^\circ$ to 0.17 at $i=85^\circ$, whereas for $a=0.5$ it increases from 0.055 at $i=0^\circ$ to 0.09 at $i=85^\circ$, with a minimum of 0.042 at $i=45^\circ$.


\subsection{Parameter Space}
\begin{figure}
    \centering
    \includegraphics[width=0.45\textwidth]{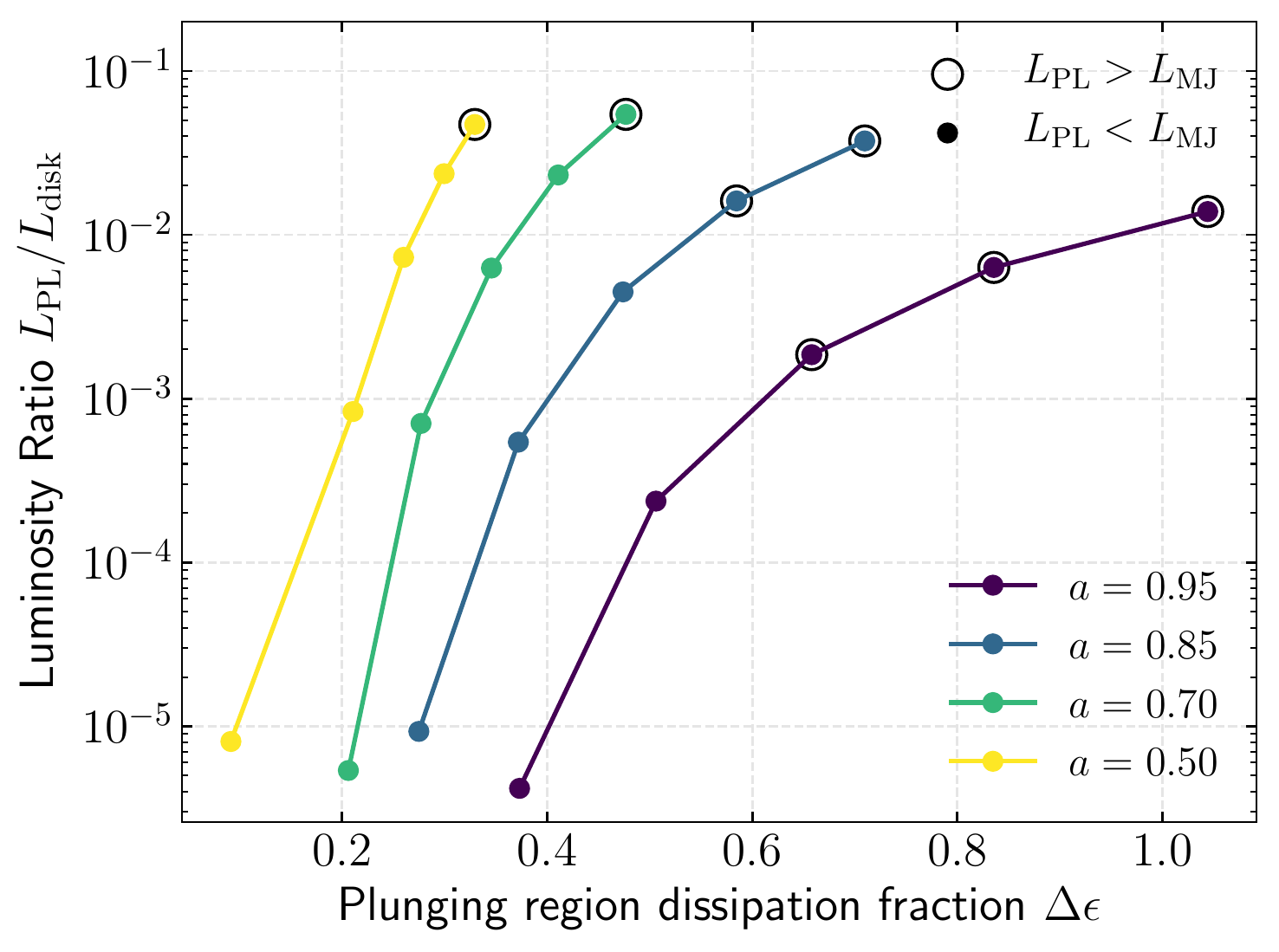}
    \caption{Frequency-integrated luminosity in X-rays ($1~{\rm keV}<\nu<1~{\rm MeV}$) in predicted power law compared to the thermal blackbody disk for different model parameters, assuming an inclination angle $i=60^\circ$. Models where nonthermal electron emission dominates over thermal electron emission, satisfying $L_{\rm PL}>L_{\rm MJ}$, are marked with large black circles.}
    \label{fig:PLdisk}
\end{figure}

Figure~\ref{fig:PLdisk} shows the power law to disk luminosity fraction as a function of plunging region dissipation and black hole spin. For $i=60^\circ$, the plunging region can emit up to 1\% of the accretion disk luminosity for a high-spin black hole, and even higher for lower spin black holes. At constant spin, disks with less magnetic torque at the ISCO and hence a lower accretion efficiency $\Delta\epsilon$ have less nonthermal luminosity because weaker magnetic fields lead to less efficient cooling of high-energy particles, whereas the thin disk emission is constant with $\Delta\epsilon$. At fixed $\Delta\epsilon$, lower black hole spins lead to a more visible power law because the plunging region extends to larger radii, increasing the magnitude of the nonthermal plunging region luminosity and decreasing the disk luminosity. For $\Delta\epsilon\approx0.4$, the power law's X-ray luminosity increases by a factor of 400, from $10^{33}$ erg/s for $a=0.95$ to $4\times10^{35}$ erg/s for $a=0.70$. This increase is due to a combination of a larger area radiating 90\% of the emission and redshifts closer to unity for smaller spins. The thin disk's X-ray luminosity decreases for decreasing spin by a factor of ten, from $10^{38}$ to $10^{37}$ erg/s. These combined effects explain why $L_{\rm PL}/L_{\rm disk}$ is a factor of 4000 different for the same magnetization.

Our model demonstrates two regimes for the plunging region X-ray emission: one where nonthermal electrons dominate and one where thermal electrons dominate. For fixed spin, weaker magnetic fields lead to less cooling by nonthermal electrons and thus require more thermal cooling and higher thermal temperatures to balance the total heating rate. Higher thermal temperatures and a less prominent power-law lead to a significant decrease in $L_{\rm PL}/L_{\rm MJ}$ with decreasing accretion efficiency. Nonthermal electrons dominate only at the highest disk magnetizations because particle acceleration requires strong magnetic fields. For less strongly magnetized disks, thermal electrons dominate the plunging region emission; in this regime, we might expect the observed power law to come from thermal electrons with a rapidly-increasing temperature rather than nonthermal electrons~\citep{zhu2012}. Figure~\ref{fig:PLdisk} marks the regime where nonthermal electrons dominate with black circles around the colored markers. 

\begin{figure}
    \centering
    \includegraphics[width=0.45\textwidth]{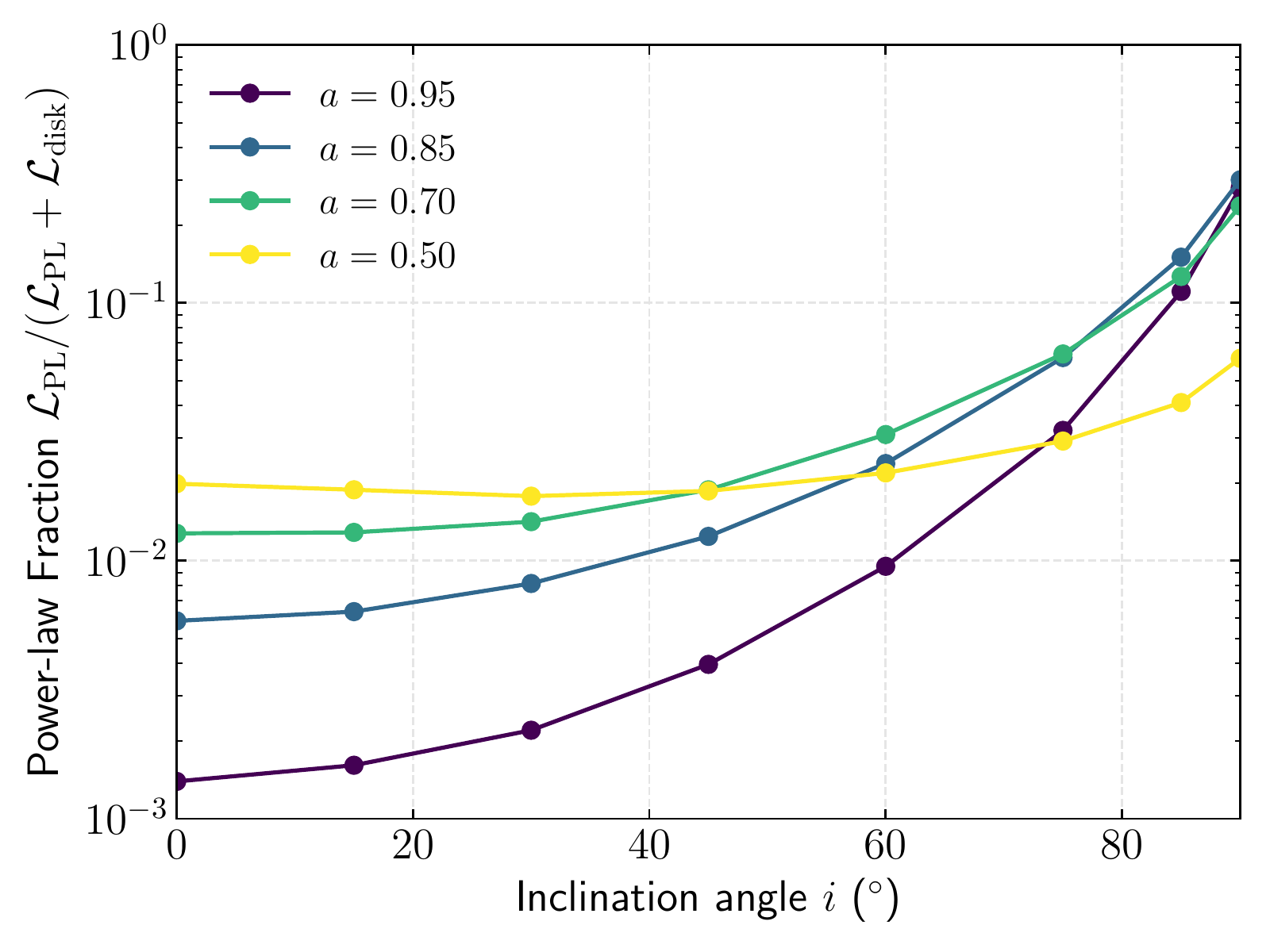}
    \caption{Power-law fraction (Equation~\ref{eq:plf}) as a function of inclination angle. Calculated from the most strongly magnetized plunging region parameters such that nonthermal plunging region emission dominates thermal plunging region emission.}
    \label{fig:PLF}
\end{figure}

To facilitate comparison with observations, we show our model's predictions for the power-law fraction (PLF) as a function of inclination angle and spin in Figure~\ref{fig:PLF}. The PLF is defined as 
\begin{equation}
    {\rm PLF}=\frac{\mathcal{L}_{\rm PL}}{\mathcal{L}_{\rm PL}+\mathcal{L}_{\rm disk}}\label{eq:plf}
\end{equation}
where $\mathcal{L}_{\rm PL}=L_{\rm PL}({\rm 1~keV}<\nu<{\rm 100~keV})$ is the power-law luminosity in the X-rays and $\mathcal{L}_{\rm disk}=L_{\rm disk}({\rm 0.001~keV}<\nu<{\rm 100 ~keV})$ is the disk luminosity over a wide range of frequencies, to avoid cutting off low-spin blackbody emission that peaks at $\sim$1 keV ~\citep{dunn2010,munozdarias2013}. For the high-spin case, the PLF is almost the same as $L_{\rm PL}/L_{\rm disk}({\rm 1~keV}<\nu<{\rm 1~MeV})$ because $L_{\rm PL}\ll L_{\rm disk}$. For lower spins, the PLF is lower than $L_{\rm PL}/L_{\rm disk}({\rm 1~keV}<\nu<{\rm 1~MeV})$ by a factor of 2 - 3 because of the change in denominator. The parameters in Figure~\ref{fig:PLF} are the most strongly magnetized disks where the nonthermal emission dominates thermal emission from the plunging region; for the least magnetized cases, the PLF is less than $10^{-3}$ (not shown) and thermal plunging region electrons would significantly change the shape of the emission above 10 keV. Overall, the PLF calculated from the model lies between $10^{-3}$ and 0.3, values consistent with observations of the soft state~\citep{dunn2010}.

\section{Discussion}\label{sec:discussion}
\subsection{Application to Astrophysical Systems}
Our model broadly agrees with observations of X-ray binaries in the soft state. The soft state usually exhibits a photon index $\Gamma\gtrsim2$~\citep{remillard2006}, corresponding to model fits with an electron power-law index $p\gtrsim3$. The model's most strongly magnetized parameters give an electron power-law index of ~$3-4$ (Figure~\ref{fig:prescription_quantities}), which increases for decreasing magnetization. The soft state's gamma-ray tail has luminosities on the order of $10^{35}-10^{37}$ erg/s above 50 keV~\citep{grove1998}. Our model finds power-law luminosities above 50 keV between $10^{34}-10^{37}$ erg/s, within the same order of magnitude. Calculations from observations find that the PLF varies between $10^{-3}$ and ~0.2 in the soft state of XRBs~\citep{dunn2010}, which agrees well with the values found from our model (Figure~\ref{fig:PLF}). As yet, trends of the PLF with inclination angle and spin cannot be found from observations~\citep{munozdarias2013}. Testing our predictions concretely against observations will require less uncertainty in the PLF. 

The power-law emission in our model comes from synchrotron emission, whereas typical models for the power law assume inverse Compton scattering of blackbody disk photons off nonthermal particles~\citep{mcconnell2002,gierlinski1999}. Our model therefore predicts more strongly polarized emission for the soft state's emission above ~10 keV than other models where inverse Compton scattering dominates. Current polarization measurements set an upper limit of 70\% for Cyg X-1's 0.4 - 2 MeV emission in the soft state~\citep{rodriguez2015}, which does not constrain the radiation mechanism.

The hard state of XRBs such as Cyg X-1 shows an excess above 400 keV that fits well to a hybrid thermal/nonthermal model with a power-law electron distribution index between 3.5 and 5~\citep{mcconnell2000}. The nonthermal electrons in that model could also be accelerated by turbulence or magnetic reconnection. If the hierarchy of timescales satisfies Equation~\ref{eq:hierarchy}, the multizone equilibrium model presented in Section~\ref{sec:methods} could be adapted to a geometry relevant to the hard state. 

\subsection{Model Limitations}
Many of our model assumptions were motivated by GRMHD simulations. Our finding that ions and electrons decouple in the plunging region is consistent with estimates of the thermalization time from single-temperature, ideal GRMHD simulations~\citep{zhu2012}. In fact, the decoupling region could extend even outside the ISCO. The~\citet{gammie1999} model artificially sets the radial velocity to zero at the ISCO, forcing a thin disk at $r>r_{\rm ISCO}$ whereas the thin disk could in principle thicken at larger radii depending on the dominance of Coulomb collisions. For simplicity, in our model, the ion temperature and thus the ratio $H/R$ was discontinuous at the ISCO. This discontinuity is not present in full 3D GRMHD simulations; instead, a transition region forms between a thin disk and the hot flow~\citep{hogg2018,liska2022}. Such a transition region would likely result in a less distinct bump from the thermal electrons in the plunging region. GRMHD simulations also motivate magnetic reconnection as an important dissipation mechanism, demonstrating hallmarks of reconnection such as current sheets and plasmoids throughout a magnetically-dominated disk~\citep{ripperda2022,scepi2022}. Our prescription for $\gamma_2$ and $p$ does not include the impact of particle anisotropy or magnetic guide field on the electron distribution~\citep{comisso2020,werner2021}, which could alter the observed spectrum. Including a guide field tends to steepen $p$ and decrease $\gamma_2$ for a pair plasma~\citep{werner2017,werner2021}. Pitch-angle anisotropy in a pair plasma can also lead to temperature anisotropy for the thermal bulk~\citep{comisso2019,comisso2020}, which could affect radiation from the thermal plunging region electrons. Unlike GRMHD simulations, our model does not include vertical structure and we assumed a constant $H/R=1$ and a uniform distribution of the dissipation throughout the plunging region volume. Vertical structure would presumably lead to a dense equatorial region with more optically thin upper parts of the plunging region. We also assume uniform dissipation such that $Q_+ \propto \rho$, as commonly done in disk atmosphere models \citep{davis2006}.

For simplicity, we did not include full radiative transport in our model. Instead, we assumed that the thermal plunging region electrons cooled solely through thermal bremsstrahlung, neglecting thermal synchrotron radiation because the average Lorentz factor of the thermal particles was less than 2 (i.e. temperatures less than $\sim10^9$ K). We also neglect synchrotron self-absorption, since this effect is only important for frequencies below $\sim10^{15}$ Hz~\citep{wardzinski2000}, outside the range we consider. We neglect cyclotron radiation for the same reason. We restrict the observed radiation from the nonthermal electrons to be either from synchrotron radiation or from inverse Comptonization of soft disk photons, i.e. we neglect synchrotron self-Compton emission (SSC). Neglecting SSC does not result in large errors for small optical depth and spectral index $>1$, where synchrotron emission dominates over SSC~\citep{wardzinski2001erratum}. We also do not consider synchrotron and inverse Compton processes simultaneously; instead, we assume one or the other dominates. We also assume a simple thermal + power-law electron distribution function, whereas Fokker-Planck models evolving a power-law electron distribution thermalizing under electron-electron collisions show a more complicated structure~\citep{nayakshin1998}. Although pair production could become important in this regime, it remains unclear how pair-regulated reconnection would change when synchrotron losses are included~\citep{mehlhaff2021}. We note that Compton reflection off the cold disk could increase the luminosity above 10 keV by order unity for sandwich-type models~\citep{zdziarski2004}. Including Compton reflection of the plunging region radiation off the cold disk could also increase the high-energy luminosity and lead to higher power-law fractions. More detailed modelling of reflection features such as the Fe K$\alpha$ emission line is beyond the scope of this work.

\section{Conclusions}\label{sec:conclusions}
We have presented a semi-analytic model for the soft state of XRBs where the high-energy power-law tail is produced by nonthermal electrons in the plunging region. We demonstrated the feasibility of having nonthermal electrons in the plunging region by examining the hierarchy between electron-proton thermalization time, electron-electron collision time, electron cooling time, electron acceleration time, and the infall time (Section~\ref{sec:dynamical}, Figure~\ref{fig:timescales}). Using an analytic dynamical background and the results of particle-in-cell simulations of magnetic reconnection, we constructed a steady-state model for the electron distribution function at each radius in the plunging region (Section~\ref{sec:methods}). The nonthermal electrons in this model produce an observable power law from $10 - 1000$ keV with a photon spectral index $\Gamma\gtrsim2$ (Figure~\ref{fig:exspectrum}). By exploring the model parameter space, we show that plunging region emission from nonthermal electrons dominates over thermal electrons for strongly-magnetized models and vice-versa for less strongly-magnetized models, suggesting a~\citet{zhu2012}-type model. We found fractions of power law to total X-ray luminosity (power-law fraction, PLF) consistent with observations for all values of spin (Figure~\ref{fig:PLdisk}). We predict an increase in the observed PLF with inclination angle (Figure~\ref{fig:PLF}). Although the PLF in our model is consistent with observational values, testing the trend with inclination angle and spin will require more observations.

Future work using numerical simulations could explore how and where the decoupling happens, as well as its possible role in spectral state transitions and jet/wind launching.

\section*{Acknowledgements}
The authors thank Mitch Begelman, Dmitri Uzdensky, and the referee, Luca Comisso, for helpful comments. AMH acknowledges support from the National Science Foundation Graduate Research Fellowship Program under Grant No. DGE 1650115. This work was supported in part by NASA Astrophysics Theory Program grants NNX16AI40G, NNX17AK55G, and 80NSSC20K0527 and by an Alfred P. Sloan Research Fellowship (JD).

\section*{Data Availability}
The simulation data underlying this article will be shared on reasonable request to the corresponding author.



\bibliographystyle{mnras}
\bibliography{refs} 

\bsp	
\label{lastpage}
\end{document}